# Automated Testing of COBOL to Java Transformation

Sandeep Hans
*IBM Research*
shans001@in.ibm.com

Atul Kumar
*IBM Research*
kumar.atul@in.ibm.com

Toshikai Yasue
*IBM Research*
yasue@jp.ibm.com

Kouichi Ono
*IBM Research*
onono@jp.ibm.com

Saravanan Krishnan
*IBM Research*
sarkris5@in.ibm.com

Devika Sondhi
*IBM Research*
devika.sondhi@ibm.com

Fumiko Satoh
*IBM Research*
sfumiko@jp.ibm.com

Gerald Mitchell
*IBM Z Software*
gerald.mitchell@ibm.com

Sachin Kumar
*IBM Z Software*
sachinkumar@ibm.com

Diptikalyan Saha
*IBM Research*
diptsaha@in.ibm.com

## ABSTRACT

Recent advances in Large Language Model (LLM) based Generative AI techniques have made it feasible to translate enterprise-level code from legacy languages such as COBOL to modern languages such as Java or Python. While the results of LLM-based automatic transformation are encouraging, the resulting code cannot be trusted to correctly translate the original code, making manual validation of translated Java code from COBOL a necessary but time-consuming and labor-intensive process. In this paper, we share our experience of developing a testing framework for IBM Watsonx Code Assistant for Z (WCA4Z) [5], an industrial tool designed for COBOL to Java translation. The framework automates the process of testing the functional equivalence of the translated Java code against the original COBOL programs in an industry context. Our framework uses symbolic execution to generate unit tests for COBOL, mocking external calls and transforming them into JUnit tests to validate semantic equivalence with translated Java. The results not only help identify and repair any detected discrepancies but also provide feedback to improve the AI model.

## 1 INTRODUCTION

Maintaining legacy applications is a growing challenge due to the scarcity of skilled professionals [19]. These applications are often written in outdated languages like COBOL and PL/I, utilizing legacy databases such as IMS (Information Management System) and transaction processing middleware like CICS (Customer Information Control System). Migrating these applications to modern infrastructures, such as cloud environments, presents significant difficulties. As a result, transforming legacy applications to contemporary programming languages and platforms has become a crucial issue for many businesses. While classical program analysis techniques exist for this purpose, they often struggle with scalability and produce code that is difficult to understand and maintain. Recent advancements in Large Language Model (LLM) based translation tools [5] offer the potential to generate more readable code. However, these tools face challenges with accuracy, especially for legacy languages. This necessitates robust validation of the transformation.

There are many aspects of transformation validation such as functional validation which ensures input-output equivalence between the original and transformed counterpart, and other nonfunctional criteria such as code quality, readability, performance, security, etc. Functional validation is particularly challenging and time-consuming due to the scarcity of developers who know both COBOL and Java languages and the relations between them. Therefore, this paper presents an automated framework and methodology to test the functional equivalence of legacy COBOL code and its modern Java counterpart, reducing the manual burden.

Our framework automatically generates test inputs for COBOL, executes them to gather output, and produces JUnit tests to verify input-output equivalence. While our test-based approach cannot guarantee functional equivalence due to practical constraints like limited path and data space exploration, it identifies translation issues when tests fail and increases confidence in the translation when they pass. Additionally, we incorporate static analysis to detect equivalent resource-interacting statements between COBOL and Java, reporting missing or incorrect equivalence. To address failures, our framework leverages generative AI to suggest automatic code repairs.

A key challenge our solution addresses is resource unavailability during test execution. Translation validation teams often lack access to all required resources, such as databases and file systems, or face difficulties in setting them up—particularly when navigating multiple program paths. Our approach mitigates these issues through mocking, making the validation process more practical and efficient.

Early work in validating code transformation for embedded applications, such as by Van Engelen et al. [42], focuses on equivalence checking for small changes using program representation. More recent work by Zhang et al. [45] presents a unit testing-based solution to filter out invalid translations between languages such as Java to Python and Python to C++, leveraging automatic unit test generation tools like EvoSuite. Various works address parts of the problem we tackle; for instance, Cobol-Check [1] functions as a unit testing framework for COBOL. There have been prior efforts in test data generation for COBOL using symbolic execution, like those by Sasaki et al. [34] and Iwama et al. [31].

Existing techniques for validating code transformation, while insightful, have significant drawbacks when applied to large-scale legacy systems like COBOL. Approaches such as those by Van Engelen et al. [42] and Zhang et al. [45] focus on small-scale changes or simpler language transformations but lack support for automated test case generation and fail to handle external resources or enterprise-level constructs such as CICS and Db2. Frameworks like Cobol-Check [1] require resource setups that are often impractical in typical development environments, and symbolic execution methods do not address the complexities of enterprise COBOL programs. Additionally, while Java mocking tools like Mockito [7] and EasyMock [3] exist, JUnit generation frameworks lack the capability to generate resource mockings. These limitations make existing methods unsuitable for complex, large-scale applications. Our approach bridges these gaps by automating the generation and execution of unit tests, including resource mocking, to ensure accurate and scalable validation of translated code.

Our solution generates unit tests for COBOL by considering program and resource input variables through static symbolic execution which accurately models enterprise COBOL semantics with byte-level accuracy ensuring branch coverage. We automate the execution of COBOL unit tests on a mainframe system with appropriate mocking of external resources, enabling thorough testing even in development environments that lack these resources. Additionally, using the outputs from COBOL test executions, we automatically generate JUnit test cases with equivalent resource mocking, generating assertions based on COBOL test outputs and variable mappings to ensure accurate testing. By adopting a pragmatic approach based on branch coverage, we ensure comprehensive testing, while failure test cases provide insights into points of failure to aid in the transformation process. Aspects such as performance and security are beyond the scope of this work.

Our work is driven by questions on the accuracy and performance of our methodology in generating unit tests for COBOL applications, the coverage achieved by the automatically generated tests, and the accuracy of the translated Java unit tests in reflecting the original COBOL behavior. To answer these questions, we conduct experiments, including evaluating the coverage of the test generator and its impact on test accuracy and assessing the effectiveness of the resource mapping component in handling external resource calls.

This paper addresses the critical industrial problem of testing the equivalence between original COBOL and translated Java code through a novel methodology. To the best of our knowledge, we are the first to develop a comprehensive framework for testing enterprise COBOL-to-Java code translation. The paper offers the following contributions:

- We have developed a static symbolic execution algorithm for modeling the full semantics of enterprise-level COBOL code by accurately modeling the data type and statement semantics. This approach accurately generates both program and resource inputs, removing the dependency on external systems during testing.
- We leverage COBOL debugger-based program execution, which allows precise control over program points and dynamic data inclusion at breakpoints. Through the use of debugger stepexecution data substitution and exclusion, we enable mocking and controlled execution without relying on external resources, ensuring accurate simulation of the program's behavior.
- We check for functional equivalence between the original COBOL and translated Java code by automatically generating JUnit tests

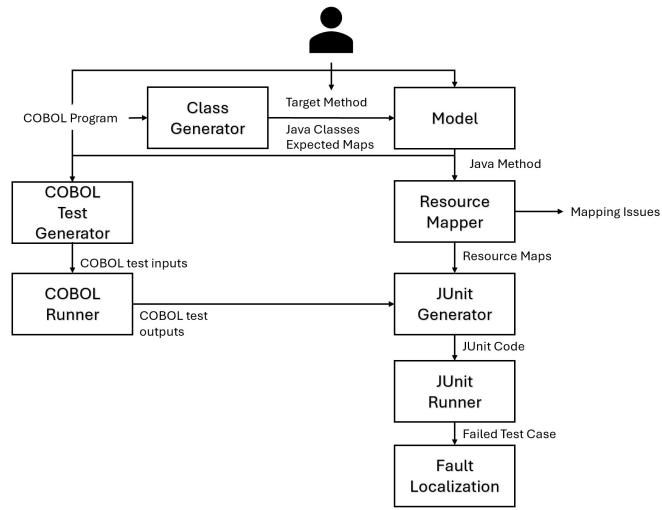

Figure 1: Architecture

with equivalent resource mappings and mocking mechanisms, which maintain consistency across both languages.
- In our experimental evaluation, we tested 33 COBOL programs from Genapp[4] and IMS dataset, consisting of 105 paragraphs, 51 of which contained multiple execution paths. Our results show that 43 paragraphs (84%) achieved full branch coverage, while 46 (90%) exceeded 80% coverage. We also evaluated 54,284 COBOL programs from an internal IBM Compiler benchmark dataset, which include 94,524 paragraphs covering the entire range of COBOL language statements, their variants, and data definitions. Among these, 87,725 paragraphs (92.8%) achieved full branch coverage. Furthermore, we evaluated 936 COBOL programs from the StarCoder [33] dataset, which include 2,871 paragraphs, out of which 2,707 paragraphs achieved full branch coverage. Additionally, the methodology significantly reduced manual effort and improved the efficiency of code translation validation, providing substantial time savings and a more streamlined experience.

## 2 APPROACH

### 2.1 Architecture

The tool's architecture is presented in Figure 1. The system takes a COBOL program as input, typically representing a specific functionality within a COBOL application. The first component, known as the class generator, analyzes the various record structures and paragraphs within the COBOL program to generate a set of Java classes, member variables, and method signatures. It determines how each COBOL variable will be translated into Java—whether as a local variable, method parameter, or class variable—and creates a mapping (referred to as *CJMap*) between COBOL and Java elements, such as records to classes, paragraphs to functions, and COBOL variables to different forms of Java variables. This mapping, along with a user-selected COBOL paragraph, is then used by the generative AI model to generate a corresponding Java function body that integrates seamlessly with the rest of the classes, including necessary library imports. An example Java transformation is provided in Figure 3 for the COBOL code in Figure 2. The details

```
1       INSERT-CUSTOMER.
2       MOVE ' INSERT CUSTOMER' TO EM-SQLREQ
3       IF LGAC-NCS = 'ON'
4       EXEC SQL
5       INSERT INTO CUSTOMER( CUSTOMERNUMBER, FIRSTNAME, LASTNAME)
6       VALUES (:DB2-CUSTOMERNUM-INT, :CA-FIRST-NAME, :CA-LAST-NAME)
7       END-EXEC
8       IF SQLCODE NOT EQUAL 0
9       MOVE '90' TO CA-RETURN-CODE
10      PERFORM WRITE-ERROR-MESSAGE
11      EXEC CICS RETURN END-EXEC
12      END-IF
13      ELSE
14      EXEC SQL
15      INSERT INTO CUSTOMER ( CUSTOMERNUMBER, FIRSTNAME, LASTNAME )
16      VALUES ( DEFAULT, :CA-FIRST-NAME,:CA-LAST-NAME)
17      END-EXEC
18      IF SQLCODE NOT EQUAL 0
19      MOVE '90' TO CA-RETURN-CODE
20      PERFORM WRITE-ERROR-MESSAGE
21      EXEC CICS RETURN END-EXEC
22      END-IF
23      *        get value of assigned customer number
24      EXEC SQL
25      SET :DB2-CUSTOMERNUM-INT = IDENTITY_VAL_LOCAL()
26      END-EXEC 27 END-IF.
28       MOVE DB2-CUSTOMERNUM-INT TO CA-CUSTOMER-NUM. 29      EXIT.
```

Figure 2: COBOL paragraph

of the class generator and the model are beyond the scope of this paper and will not be further elaborated upon.

The remaining components are designed to validate the functional equivalence between the COBOL paragraph and the compilable Java function by generating and executing test cases with appropriate mocking. The COBOL test generator performs a static symbolic execution of the COBOL program to generate input values for variables and compile a list of output variables (excluding their values). The COBOL test runner then uses these inputs to execute the program without interacting with external resources using a customized debugger-based execution, collecting the output variable values. The resource mapper determines the mapping (*CJResourceMap*) between COBOL external call arguments and Java call arguments, enabling equivalent mocking between the two units. The JUnit generator component creates JUnit code that initializes the Java inputs using the COBOL inputs and various maps - *CJMap* and *CJResourceMap*. It also generates assertions for the Java output variables, comparing their values to the corresponding COBOL output variable values.

## 2.2 COBOL Test Input Generation

Our primary objective is to develop a *scalable* and *accurate* solution capable of generating test inputs that maximize path coverage in COBOL paragraphs, thereby enabling equivalence checking across these paths. The inputs should comprise both *program inputs* (used before being defined) and *resource inputs* (variables provided by external calls). Resource inputs are critical due to *resource availability* constraints. Additionally, for each test case, the solution must accurately identify all program output (unused definition) and resource output variables (variables whose values are going out of the paragraph through resource statements) along each path, ensuring output values can be captured during execution.

For instance, in the INSERT-CUSTOMER paragraph (depicted in Figure 2), which contains four paths, the path traversing line

```
1                       public void
                        mainlineInsertCustomer(long
                        db2CustomernumInt,
2                       String lgacNcs) {
3                       ErrorMsg errorMsg = new ErrorMsg();
4                       errorMsg.getEmVariable().setEmSqlreq("
                        INSERT CUSTOMER");
5
6                if (lgacNcs.equals("ON")) {
7                try {
8                String sql = "INSERT INTO CUSTOMER
9                (CUSTOMERNUMBER, FIRSTNAME, LASTNAME)
10               values (?, ?, ?)";
11               PreparedStatement ps = DriverManager
12               .getConnection("conn")
13               .prepareStatement(sql);
14               ps.setLong(1,db2CustomernumInt);
15               ps.setString(2,this.genasa1Customer.getFirstn
                 ame());
16               ps.setString(3,this.genasa1Customer.getLastna
                 me());
17               ps.executeUpdate();
18               ps.close();
19      }
20          catch(SQLException exception) {
21          this.caReturnCode = 90;
22          errorMsg.mainlineWriteErrorMessage(dfhcommarea1);
23          return;
24          }
25          }
26               else {
27               try {
28               String sql = "INSERT INTO CUSTOMER
29               (FIRSTNAME, LASTNAME) values (?, ?)";
30               PreparedStatement ps = DriverManager
31               .getConnection("conn") 32   .prepareStatement
33               (sql, Statement.RETURN_GENERATED_KEYS);
34               ps.setString(1,this.genasa1Customer.getFirstname());
35               ps.setString(2,this.genasa1Customer.getLastname());
36               ps.executeUpdate();
37               ps.close();
38               ResultSet rs = DriverManager.getConnection("conn")
39               .createStatement().executeQuery
40               ("SELECT IDENTITY_VAL_LOCAL() FROM
                 SYSIBM.SYSDUMMY1");
41               rs.next();
42               db2CustomernumInt = rs.getLong(1);
43      }
44          catch(SQLException exception) {
45          this.caReturnCode = 90;
46          errorMsg.mainlineWriteErrorMessage(dfhcommarea1);
47          return;
48          } 49       }
```

Figure 3: Translated Java function

numbers (2-8, 12, 28, 29) includes program inputs {LGAC-NCS, DB2-CUSTOMERNUM-INT, CA-FIRST-NAME, CA-LAST-NAME}, resource inputs {SQLCODE (line 5)}, program outputs {EM-SQLREQ, CA-CUSTOMER-NUM, SQLCODE}, and resource outputs {DB2CUSTOMERNUM-INT (line 5), CA-FIRST-NAME (line 5), CA-LASTNAME (line 5)}.

To address these requirements, we selected a static symbolic execution (SSE) [18] approach due to its ability to deterministically explore program paths, compute resource inputs and outputs for each path, and present the equivalence-checking path to the user.

In SSE, a set of execution paths is determined based on a defined path selection strategy. For each path, a constraint logic formula is constructed via symbolic execution, called path constraint, reflecting the semantics of the statements on input necessary to execute that path. Th path constraint is then solved using a constraint solver (Z3 [11], in our case) to generate satisfiable solutions.

While SSE offers significant advantages, it has known limitations: scalability challenges due to path explosion [16], accuracy issues stemming from the constraint solver's expressiveness, and implementation overheads requiring precise semantic modeling of the target language. Below, we describe our strategies and experiences in addressing these challenges to create an enterprise-grade test generation solution for COBOL.

*2.2.1 Coverage Criteria.* Given the computational expense of allpath exploration, we adopt branch coverage as our coverage criterion. This approach balances scalability and bug-finding ability. Our solution systematically covers unexplored branches during depth-first traversal of the control flow graph in SSE, akin to the approach described in [43]. As demonstrated in Section 3, this strategy achieves high coverage of instructions.

*2.2.2 Data Representation.* COBOL provides considerable flexibility in data representation, particularly for numerical data, which may be stored as integers, decimals, or floating-point values in formats such as binary, Zoned Decimal, Packed Decimal, and single or double precision [21]. The PICTURE clause further allows precise declaration of data item formats [21]. Additionally, the REDEFINES clause permits interchangeable use of string and numeric data without explicit conversion [22].

To ensure accurate internal data modeling, we employ a bytelevel representation. Each data item is uniformly represented as a sequence of bytes, and every byte is treated as an 8-bit BitVector within the Z3 constraint language. Based on the datatype, the appropriate range constraint with the byte is associated. This is referred to as type constraints. For example, for COBOL declaration '01 TEMPERATURE PICTURE 999V99' 5 bytes are allocated, each byte ranging from 0xF0 to 0xF9. When the variable is involved in conditional branches (say $TEMPERATURE > 98.6$) then the variable value is derived by adding the place value expressions for individual bytes.

This approach accommodates COBOL's flexibility while maintaining accuracy in constraint formulation. Such modeling raises the obvious question on the scalability of the technique, as the number of constraints is typically high. Two observations come to our advantage: 1) the size of a COBOL paragraph is typically small (typically less than 500 lines), and 2) the number of memory locations involved in path constraints is typically less. In Section 3, we therefore show the efficiency of our technique.

*2.2.3 Statement Semantics Modeling.* Accurate input-output semantic modeling of statements is essential for SSE. COBOL's syntax encompasses over 70 statement types and more than 1300 grammar rules, underscoring its syntactic complexity. To address this, we developed an Intermediate Representation (IR) with data types closely aligned to COBOL with 55 statement types. COBOL Abstract Syntax Trees (ASTs) are translated into the IR, and symbolic execution is implemented on top of this representation. To our knowledge, ours is the first solution to handle such a comprehensive range of COBOL syntactic variations in symbolic execution.

To ensure semantic accuracy, we employed few verification techniques. For instance, we verified that input generated from path constraints accurately traversed the intended path using an interpreter (written by us) during the development phase. Path conformance was further validated at the integration testing phase using debugger-based execution, with discrepancies flagged for refinement. The IBM Z Cobol compiler benchmark, encompassing all COBOL syntax examples, was used for unit testing the SSE.

```
{
  "Path": [
    2, 3, 4, 5, 7, 8, 12, 28, 29
  ],
  "Program-Inputs": {
    "LGAC-NCS": "ON",
    "DB2-CUSTOMERNUM-INT": 1000,
    "CA-FIRST-NAME": "ABC",
    "CA-LAST-NAME": "DEF"
  },
  "Resource-inputs": [
    {"variable": "SQLCODE", "line": 5, "value": 0}
  ],
  "Program-outputs": [ "EM-
    SQLREQ",
    "CA-CUSTOMER-NUM",
    "SQLCODE"
  ],
  "Resource-outputs": [
    {"variable": "DB2-CUSTOMERNUM-INT", "line": 5},
    {"variable": "CA-FIRST-NAME", "line": 5},
    {"variable": "CA-LAST-NAME", "line": 5}
  ]
}
```

Figure 4: Generated COBOL test case

For enterprise COBOL constructs like CICS [24] and IMS [25], which involve resource statements, we created a catalog of use-def variables for each syntactic variation in statements. This enables precise identification of resource inputs and outputs. For example, the input and output arguments for Get-Unique statement from IMS Database [23] is specified by the "arg_type" parameter in the below JSON.

```
{
    "called_prog_code" : "CBLTDLI",
    "function_code" : "GU",
    "description" : "get unique",
    "arguments" :
    [
      {
        "arg_position":0,
        "arg_description": "database PCB must for interactions with IMS
          DB",
        "arg_type": "output"
```

```
      },
      {
        "arg_position":1,
        "arg_description": "segment I/O area",
        "arg_type": "output"
      },
      {
        "arg_position":2,
        "arg_description": "segment search argument",
        "arg_type": "input",
        "is_optional":"true",
        "is_multi_valued":"true"
      }
    ]
  }
```

*2.2.4 Constraint Solving.* Till now we have explained two types of constraints: (1) type constraints, reflecting data-type specifications within the byte-level representation, and (2) path constraints, arising from conditional branches. Additionally, implicit constraints from statements, particularly resource statements, are incorporated. For example, in Figure 2, the SQLCODE variable assumes a value of zero for successful database operations or predefined non-zero codes otherwise [20, 26]. To enhance data realism, users can define additional regular-expression-like constraints for variables. These constraints are translated to byte-level constraints and are added to type and path constraints whenever available.

A sample output of the COBOL test generator is shown in Figure 4. It contains the values for the program input variables and resource input variables, necessary for traversing the shown path corresponding to the COBOL paragraph shown in Figure 2.

We believe this represents a compelling industry use case for the widely recognized symbolic execution algorithm. Our approach not only addresses the challenges posed by the scale of industrial COBOL but also effectively manages practical constraints, such as resource limitations.

## 2.3 Debugger-based Execution

COBOL Debugger-based program execution allows for precise manipulation of the program points. COBOL debugger allows an initial controlled setup to focus on sub-unit execution. With dynamic data inclusion at breakpoints, the COBOL debugger enables mocking through the use of debug step execution data substitution and exclusion in statement step-over. This allows for the application of generated data to the program within the conditions necessary for correct output values to be created and recorded correlating to the program execution pattern of the data without COBOL external calls to real systems or virtual substitution mechanisms.

For the test case shown in Figure 4, the debugger puts a breakpoint at the first statement of the path (line 2) and sets program input variables with the generated values at runtime. It puts breakpoints at the lines present in resource-input and resource-output sections. At runtime, it sets the values of resource-input variables at the corresponding statements and gathers the values of resourceoutput variables. Finally, it gathers the values of program output variables at the last statement of the path.

## 2.4 Resource Mapping

The objective of the resource mapper is to find the mapping between COBOL and Java resource statements, that perform resource operations (SQL statements, file processing statements) or make external calls (calls to other paragraphs and programs) outside the target paragraph. The mapping should include both statement and variable-level mapping to take equivalent action during the COBOL and Java mocking.

We designed the resource mapper in a way that will be easily extensible by the developer as and when the code translation LLM is extended with new COBOL constructs. The developer-friendly approach makes this design, particularly novel, creating a perfect balance between development easiness to achieve completeness and soundness.

Central to the idea is to define a language to map constructs of both languages and create a mapping between COBOL and Java. Such mapping is bootstrapped automatically and can be extended by the developer manually if required. Such a mapping will be used to create the resource mapping for the COBOL paragraph and Java function in mind. The Resource Mapper component has two steps:

- Offline Stage: It creates the map, called *CJResourceMap*, between COBOL and Java resource calls by automatically analyzing the COBOL-Java pairs used in model training.
- Online Stage: It first matches each COBOL resource call to potentially a set of Java sequences using *CJResourceMap* and provides weights representing the degree of the match. Subsequently, it solves an optimization problem that maximizes the weighted sum and performs diversified matching.

*2.4.1 The Mapping Language.* A *CJResourceMap* defines a set of mapping rules between a COBOL statement sequence pattern and a Java statement sequence pattern (usually method calls). Each statement sequence pattern has a sequence of statement types (e.g SQL-INSERT for COBOL, java.sql.Connection.prepareStatement for Java), the number of occurrences (SINGLE/MULTIPLE) of a statement type, and constraints which can be a value of a parameter/property of a statement and alias relation between the variables of different statements. Each mapping rule maps a property of a COBOL statement to an argument of a Java method call under some value constraints or custom-defined function. One such custom function is *TableFieldMatch* which correlates COBOL and Java variables in database statements equating if pointing to the same field in the corresponding database statements. An example rule is provided in Figure 5 which can map the SQL-INSERT statements in our COBOL example to the Java statement sequences.

*2.4.2 Training Data Analysis.* Writing the mapping rules covering all mapping is a tedious task. We therefore leverage the training data for the model which contains SME-written COBOL-Java equivalent snippets to bootstrap the rules. Each COBOL-Java pair already has the mapping between COBOL and Java variable mapping. Given a COBOL-Java pair containing resource statements, our analysis first identifies the mapping between COBOL and Java resource statements containing the already mapped variables. This forms mapping rules. Then it identifies the backward slice at COBOL and Java snippet to identify the subsequence of the program which is a precondition to the COBOL and Java resource statements. The alias constraints capture the slice dependencies. The rules are manually maintained and extended.

Given a COBOL program and a Java program and a set of *CJResourceMap*s, the online phase identifies resource statement mapping between them. This can be divided into two phases Mapping and Disambiguation, described below.

*2.4.3 Mapping.* In the mapping phase, for a *CJResourceMap* it identifies a set of instances of COBOL and Java sequences satisfying all the constraints. The process involves ud-chain (for value constraints) and pointer analysis (for satisfying alias constraints). Note that there can be multiple COBOL subsequences and Java subsequences matching a single rule. For example, there are two SQLINSERT statements (Line 5 and Line 15) in the COBOL program in Figure 2 and two subsequences (Lines 7-15 and Lines 25-32) mapping the SQL-INSERT *CJResourceMap* rule.

*2.4.4 Disambiguation.* This step is able to disambiguate the mapping. The disambiguation is done based on a combination of three criteria: 1) Maximal constraint match: how many mapping rules were satisfied; 2) Diversified matching: if there are two same COBOL insert statements and there are two matching Java sequences then the preferable matching should not map both COBOL statements to a single Java statement; 3) Position matching: The relative position of COBOL statements should be respected in Java as transformations typically do not change the order of conditional branches.

## 2.5  JUnit Generation

The JUnit Generator component generates JUnit test code to validate the generated Java functions corresponding to each COBOL test for the source paragraph, utilizing the COBOL inputs, outputs, *CJResourceMap* and *CJMap*. This process involves three parts: 1)

```
COBOL: insert1: (type==SQLINSERT) Java: s1:
(type==java.sql.Connection.prepareStatement, occ=SINGLE)
s2:(type==java.sql.prepareStatement.executeUpdate, occ=SINGLE) s3:
(type==java.sql.prepareStatement.(setBoolean|setByte|...), occ=MULTIPLE)
  Constraints: [alias(s1.ret,s2.obj), alias(s1.ret,s3.obj)]
MappingRule:
 TableFieldMatch(insert1, s3.argument(1),s2.argument(0))
```

Figure 5: *CJResourceMap* Rule for SQL-INSERT

creation of initialization code, where local and class-level variables are initialized using the COBOL-Java variable mapping in *CJMap* and COBOL program inputs, 2) mocking of Java resource statements using resource mapper and COBOL resource inputs, and 3) generating assertions using program and resource outputs.

Several key challenges arise in testing COBOL-to-Java translations. First, mapping COBOL data types to their Java equivalent types for initialization and assertions is complex due to differences in how each language handles data structures. Second, managing order-dependent mocking is critical, as *CJResourceMap*'s static statement-level mapping has to be elevated to the path level which can include multiple occurrences of the same statement. Third, ensuring platform-independent execution, particularly decoupling tests from the Z platform, is essential to running Java tests on different environments without dependency issues. Fourth, COBOL programs often handle implicit paths such as SQL codes differently from Java, requiring careful enforcement of these paths in testing. Lastly, managing multiple points of assertions and supporting multi-assertions within a single test case adds an additional layer of complexity.

JUnit Generator offers effective solutions to these challenges. For datatype mapping, it incorporates precise mapping rules between COBOL and Java datatypes in the assertions, ensuring that data integrity is maintained across both environments. For example, it includes checks for trimmed strings, which are common in COBOL due to fixed-length fields but need explicit handling in Java. The generator ensures that trailing spaces in COBOL strings are appropriately handled, validating that the translated Java string maintains the expected content without unnecessary padding.

Many testing techniques exist, but they often overlook external factors such as databases. Mocking is a strategy to handle the external effects of the code during unit testing. For a target program with external interactions, the generated JUnit test code mocks the behavior of these interactions using a library called Mockito. The JUnit code includes mock statements for any external interactions of the target method, which may involve database interactions (inserting or retrieving data), file system operations, or platform-specific libraries such as CICS and JZOS. This also includes generating mock implementations of Z libraries that can be used on non-Z platforms, allowing the test execution to run on any platform without requiring a Z environment.

For each test path, the JUnit generator also needs to ensure that the same resource inputs in the statement occurrence in the COBOL test case are provided to the corresponding Java variables of the Java statement occurrence. The Resource Mapper provides the statement and variable mapping. That is elevated to the statement occurrence level in the path by order-dependent mocking. For each generated mocked Java statement, the JUnit generator assigns the resource input in the same order of occurrence as the corresponding COBOL statement in the COBOL test path.

To enforce implicit paths such as SQL codes, JUnit tests incorporate logic that simulates the handling of SQL error codes and other transaction paths, ensuring that the Java version behaves as expected. For example, in COBOL, specific SQL codes like −100 (no rows found) or −109 (invalid clause) control program flow, but Java throws a general *java.sql.SQLException* instead of specific codes. This difference prevents Java test cases from fully replicating COBOL's precise error handling, making SQL-specific condition validation challenging.

Finally, JUnit Generator allows for multiple assertions within a single test case, making it possible to test multiple parts of the system simultaneously, ensuring comprehensive validation without needing to break down tests into overly granular components. Through these solutions, the JUnit Generator proves to be a versatile and effective tool in addressing the challenges of validating COBOL-to-Java translations.

The algorithm begins by parsing the Java code and retrieving all variables and method invocations in the target method. It then identifies variables for mocking using the library specification and initializes their usage for the current test. A graph is created with nodes representing variables and edges representing return relationships. For each method invocation, the algorithm determines the calling variables and return type, identifies all returned variables with the same return type and earlier line numbers, and assigns edge labels based on whether the invocation line is part of the test path. Edges are added between calling and returned variables when the calling variable's line number is less. Finally, the algorithm extracts the variables used for mocking from the path in the graph. Finally, it generates assertions to match the COBOL and Java program outputs.

## 2.6 Handling Faulty Transformations

Faulty transformed Java statements may result in assertion failures or execute a different path from the expected Java path. The tool provides a fault localization mechanism by pointing out faulty statements that the user can use as reference while fixing the transformation. Given the program statements executed by a COBOL test case, the tool derives the expected Java statements ($Stmts_{exp}$) to be executed for each test case from the Resource-Mapper. The actual Java statements executed are obtained through instrumenting the
Java target method to log the statements ($Stmts_{act}$) on executing the JUnit testcases. The instrumented method-body, $Stmts_{exp}$ and $Stmts_{act}$ are fed to a large-language model instructing it to pick candidate faulty statements given the data. The response was used as a reference to fix the faults. We have experimented with the llama-3-70b-instruct model [30]. Figure 6 shows a sample prompt and the response for one of the benchmark methods. The faulty transformation here results in transforming methods parameters into local variables with the hardcoded value of `lgacNcs`. As a result, the `if` branch is traversed for any value of `lgacNcs` in the test case. As a result, the derived $Stmts_{exp}$ varies from $Stmts_{act}$. The LLM correctly points out the variable declaration statement as faulty, giving appropriate reasoning that can help the user fix the transformation.

```
1   Given this method:
2   public static void insertCustomer() {
3       CaRequest caCusReq = new CaRequest(); // Line 44
4       .
5       .
6       String lgacNcs = "ON"; // Line 47
7       if (lgacNcs.equals("ON")) {
8           .
9           .
10      }
11      else {
12          .
13      }
14  }
15
16  Expected lines to execute:
17  76, 77, 78, 79, 80, 81, 82, 83, 84, 88
18
19  Actual Executed lines:
20  50, 51, 52, 53, 54, 55, 56, 58, 59, 60, 61, 62, 63, 64
21
22  Find the line candidates that could be faulty.
23  Response:
24  The line candidates that could be faulty are:
25
26  - Line 47: String lgacNcs = "ON";
27  - Line 51: if (lgacNcs.equals("ON"))
28
29  The variable lgacNcs is always "ON" and the
30  if condition is always true, so the else
31  condition is never executed. This could be a bug in the code. The variable lgacNcs should be set based on some condition or input, not hardcoded to "ON".
```

Figure 6: Fault Localization Prompt

## 3 EXPERIMENTAL RESULTS

We seek to answer the following research questions through our experiments:

RQ1: How effectively does our methodology generate and execute unit tests for COBOL applications?

- What is the coverage achieved by the automatically generated tests in terms of program and resource inputs?

RQ2: How accurately do the translated Java unit tests reflect the original COBOL behavior?

- How consistent are the outputs between the COBOL tests and the corresponding JUnit tests?

RQ3: What is the impact of our approach on the overall process of maintaining and transforming legacy COBOL applications?

- What are the limitations and challenges encountered when applying our methodology to different COBOL applications?

### 3.1 Experimental Setup

We used programs from different applications to run the complete validation pipeline. These applications are chosen so that they cover a range of COBOL statements and also the calls to external resources such as databases and middleware/transaction frameworks. These programs are selected from the following applications:

(1) Genapp[4]: an IBM demonstration mainframe application for insurance domain. This is an interactive/online application that uses CICS for transaction processing and Db2 databases as well as VSAM files for data storage.

(2) **IMS Bank**: IMS Bank is another IBM demonstration mainframe application. This application uses IBM's IMS Database.
(3) **CompilerBenchmarkDataset**: A large internal data-set COBOL programs provided by IBM's COBOL compiler team. The dataset includes programs covering the entire range of COBOL language statements, their variants, and data definitions.
(4) **StarCoder Dataset** [33]: We used 936 COBOL programs from the StarCoder data-set that we could parse using our parser.

For each program, we selected COBOL paragraphs with maximum resource interaction for translation. For each COBOL paragraph, we manually gathered the total number of paths and branches and computed the coverage by looking at the generated test inputs. We also gathered the statistics related to generated inputs and assertions. The next subsection presents the details.

## 3.2 Results

We showcase the results of two stages (COBOL test case generation and Java JUnit generation) in Figure 7, Figure 8, Figure 9 and Table 1 with representative COBOL programs. A subset of 33 programs taken from Genapp and IMS Bank have total 105 paragraphs and 51 of them have two or more paths. Further, we consider a subset of 22 programs that include 40 paragraphs of those 51 paragraphs to showcase the results in Figure 7, 8 and 9. There are 54284 COBOL programs in the IBM COBOL Compiler Benchmark Data-set, which include 94524 paragraphs covering the entire range of COBOL language statements, their variants, and data definitions. This dataset has the following statistical distribution on the number of lines of code: minimum 3 lines, maximum 50,822 lines, mean 1612.7 and median 220. We used this data-set to automatically generate unit tests for these paragraphs using the COBOL Test Generator module.

We use 936 COBOL programs from StarCoder data-set which include 2871 paragraphs. This data-set has the following statistical distribution on the number of lines of code: minimum 3 lines, maximum 4343 lines, mean 123.79 and median 46. We used this data-set to automatically generate unit tests for these paragraphs using the COBOL Test Generator module. One test is generated (path) for most paragraphs (2352). Maximum number of tests generated for a paragraph is 16 with a mean of 1.29 and median of 1. 100 percent branch coverage is achieved for 2707 paragraphs with a mean of 97.19 percentage for all paragraphs. For 2661 paragraphs, time taken to generate tests is less than 1 second with a mean of 1.28 seconds, median of 0.14 second and maximum of 176 seconds.

*3.2.1 RQ1: How effectively does our methodology generate and execute unit tests for COBOL applications?*

Figures 7, 8 and 9 show the average branch coverage rate, total branches covered, and total paths produced per program, respectively. 43/51 paragraphs (84%) have full branch coverage and 46/51 paragraphs (90%) have above 80% branch coverage. Two possible reasons that affect the coverage are a) the presence of logical dead branch by original program implementation and b) limitations of the current heuristic approach.

For the compiler benchmark, Figure 10a shows the distribution of the number of tests generated for each paragraph. We generate one test for each path. While most paragraphs have only one path, there are paragraphs that have as many as 55 paths. Figure 10b shows the distribution of branch coverage percentage. We achieve 100 percent branch coverage for 87725 paragraphs (out of 94524). We noticed that there are some programs where the branch coverage is zero.

| Program | Paragraph | # Paths /tests | # Pro g. output | # Res. output | # Assertions | # Prog. assertions | # Res. Assertions | # Tests Pass |
|---|---|---|---|---|---|---|---|---|
| CHANN11 | FIRST-SENTENCE | 2 | 21 | 8 | 9 | 5 | 4 | 0/2 |
| KDELG | A010 | 1 | 8 | 3 | 5 | 2 | 3 | 1/1 |
| ICDBGHNP | GET-CUSTACC2 | 5 | 42 | 34 | 7 | 7 | 0 | 5/5 |
| ICDBGNP | GET-CUSTACC2 | 5 | 37 | 34 | 7 | 7 | 0 | 5/5 |
| ICGCUDAT | GET-CUSTOMER-DATA | 3 | 16 | 15 | 3 | 3 | 0 | 3/3 |
| ICLOGOUT | LOGOUT | 4 | 28 | 24 | 6 | 6 | 0 | 4/4 |
| ICSCUDAT | BEGIN | 1 | 3 | 4 | 4 | 0 | 4 | 4/4 |
| LGACDB01 | INSERT-CUSTOMER | 4 | 11 | 6 | 22 | 3 | 19 | 4/4 |
| LGAPDB01 | INSERT-POLICY | 3 | 19 | 21 | 7 | 7 | 0 | 2/2 |
| LGIPVS01 | FIRST-SENTENCE | 2 | 33 | 5 | 3 | 3 | 0 | 2/2 |

Table 1: Experimental results (RQ2) showing cobol paths and corresponding java tests.

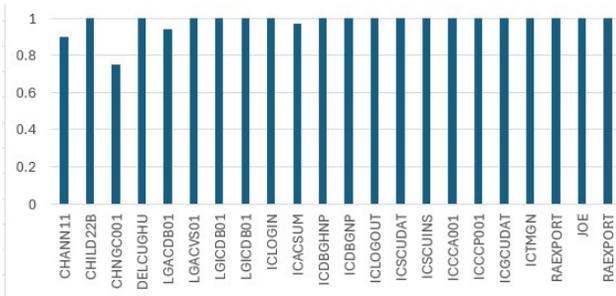

Figure 7: Branch Coverage Rate

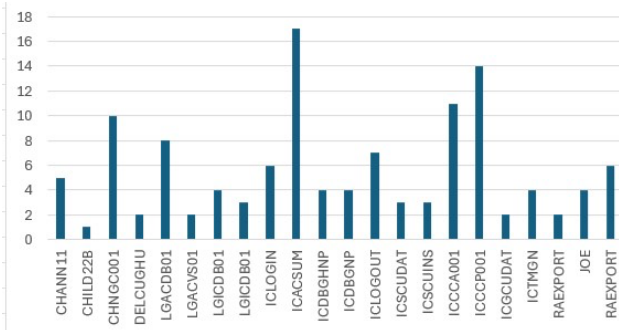

Figure 8: No of Branches Covered

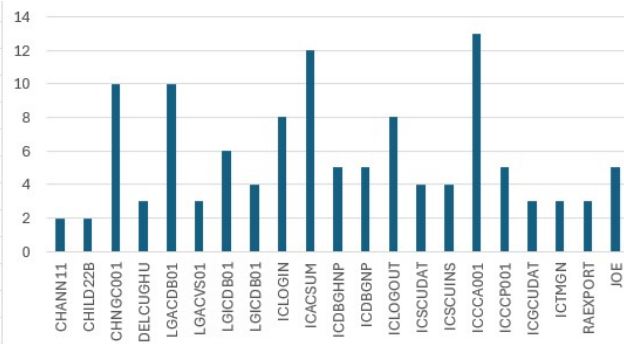

Figure 9: No Of Paths

After manually investigating the source code for those programs/paragraphs, we found that there is an unconditional GOBACK statement at the beginning of the paragraph followed by some code including conditional statements. Since this is a compiler benchmark data set, these programs are deliberately designed that way to test the compiler's behavior for emitting warnings/errors for unreachable code. There are few other programs where the branch coverage percentage is non-zero but less than 100. These programs also have unreachable code with branches. In some cases, there is conditional code inside the loops that becomes unreachable during path calculation because of the limit on the number of iterations we stop at during symbolic execution. This is a well-known problem discussed in [35].

Figure 10c shows the distribution of execution time for paragraphs. These experiments were run on a Linux Virtual Machines that has 48 cores of Intel x64 CPUs (2.2 GHz) and 196GB of total RAM. For most of the paragraphs (85657), the total time taken to generate tests is less than 1 second. In some cases, where the paragraphs are large or they have nested conditions/loops, the time taken is as high as 118 seconds. We manually checked the programs for which our system takes relatively large time to generate tests. We noticed that these programs have very large number of data items defined in arrays leading to large number of constraints. And also contain initialize statements for these arrays that gets translated to individual assignment statement in our Intermediate Representation (IR) language. For example the following code segment defining an array of 270000 nine digits numbers. And there are multiple such definitions in the same program.

```
1 JSON16n1 group-usage national.
      3 enum-largeitem pic 9(07) occurs 270000 times.
```

#### 3.2.2 RQ2: How accurately do the translated Java unit tests reflect the original COBOL behavior?

There are 4 COBOL test cases generated for 'LGACDB01' (Table 1) with 11 program output variables and 6 resource output variables. At the Java JUnit generation part, there are 22 assertions created (3 for program output variables and 19 for resource output variables). There are 4 JUnit tests created and all of them pass the tests that validates the transformation.

There are 2 COBOL test cases generated for 'CHANN11' (Table 1) with 21 program output variables and 8 resource output variables.

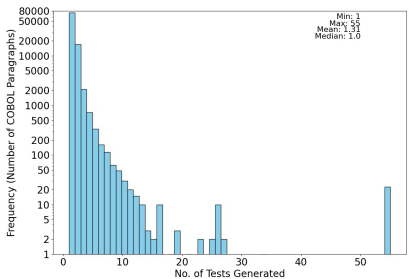
(a) #Tests for each paragraph

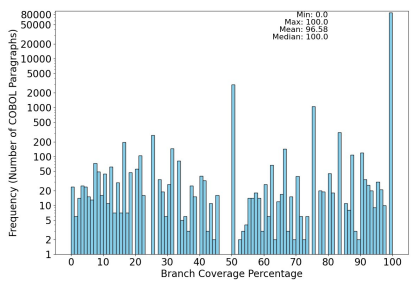
(b) Branch coverage

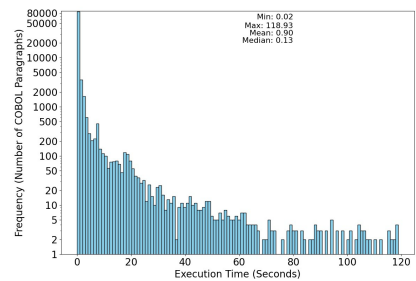
(c) Execution time

Figure 10: Experimental results for the Compiler benchmark.

At the Java JUnit generation part, there are 5 assertions created for the program variables and 4 assertions created for resource variables. There are 2 JUnit tests are created and both of them failed. On examining the values of the test input variables we could see that the following COBOL code block

```
PERFORM VARYING WS-CNT FROM 1 BY 1
    UNTIL WS-CNT > WS-LOOP-ITERATIONS OR WS-EXIT-
EARLY = 'Y'
```
has been transformed incorrectly as the Java code block below:
```
while (wsCnt <= wsLoopIterations &&
       wsExitEarly == 'N')
```

Note that the loop condition is incorrectly inverted when the condition for UNTIL in COBOL is converted to the condition for while in Java. It is correct only if the COBOL variable `WS-EXIT-EARLY` only takes values `Y` and `N`, which is not the case, and therefore test generation module can generate any value other than `Y` for the negative case. It should be: `wsCnt<=wsLoopIterations && wsExitEarly!='Y'`

Note that the number of test cases generated may not be equal to the COBOL test inputs. Consider 'ICSCUDAT' (Table 1) that has only one COBOL test case while it has 4 java test cases. All the branches of the COBOL code is part of a PERFORM block and hence one cyclic path is sufficient to capture them all.

In COBOL, specific SQL codes, such as −100 for no rows found or −109 for a missing or invalid clause, are used to handle various database conditions. These SQL codes can be checked in conditions to control the program's flow based on the type of SQL error encountered. However, Java does not have a direct equivalent for these error or non-zero SQL codes. Instead, Java uses exceptions to handle any SQL errors. Consequently, if a COBOL program checks specific SQL codes within a condition, the corresponding Java model will throw a general exception for any SQL error. This discrepancy means that specific SQL codes cannot be directly checked in Java, making it impossible to traverse the path for a particular non-zero SQL code in a Java test case. As a result, Java test cases that aim to validate specific SQL conditions from COBOL may be invalid or fail to replicate the precise error handling logic.

Also, note that the number of assertions generated may not match COBOL program outputs. This occurs because the COBOL test generation module might check a variable's value, but during transformation, the variable may become local in Java. Consequently, these local variables cannot be asserted in JUnit tests, making it difficult to replicate precise variable checks from COBOL.

The result demonstrates that our tool can successfully validate enterprise-grade COBOL transformation and able to identify potential failures in the transformation. In addition, with the scalability of the branch-coverage algorithm, we can obtain full path coverage thereby validating equivalence across all paths in the program.

*3.2.3 RQ3: What is the impact of our approach on the overall process of maintaining and transforming legacy COBOL applications?*

For some of the Genapp programs, we explored the effort saved through automated validation compared to manual validation. For instance, LGAPDB01 saw a reduction from 55 minutes to 10 minutes, while LGACVS01 decreased from 35 minutes to 8.5 minutes. Similarly, LGIPVS01 required 45 minutes manually but only 8 minutes with automated validation, and LGUCDB01 dropped from 20 minutes to 8 minutes. These findings highlight significant time savings, demonstrating the potential of automation in reducing validation efforts.

The development and evolution of our validation tool has undergone significant evolution over the past year, driven by industry feedback and practical requirements. Starting with basic COBOL constructs like MOVE, DISPLAY, and GOTO in early 2024, the tool progressively expanded to include support for SQL statements and CICS constructs, such as containers and file operations, by late 2024. These iterative updates demonstrate the tool's adaptability and emphasize the role of industry collaboration in shaping practical solutions for legacy code modernization. More details and a demo can be found in our earlier work [32].

## 4  RELATED WORK

*Software Modernization/Transformation.* COBOL, developed in 1959, remains essential for many financial and governmental institutions. Recent efforts like X-Cobol [15] provide valuable datasets for studying COBOL projects and support tools aimed at modernizing these systems. Tools such as COBREX [14] help extract business rules from COBOL using cfg-based methods, though challenges persist in handling enterprise-level programs like CICS [36].

Symbolic execution techniques, such as those proposed by Sneed [40], enhance testing accuracy for COBOL-to-Java conversions but face scalability issues. Studies highlight COBOL's resilience in transaction processing, countering misconceptions about its obsolescence [19, 29, 41].

The growing need for COBOL programmers, amid retirements, has driven modernization efforts like the migration of AS400 COBOL to Java [38, 39]. Meanwhile, tools like XMainframe, a large language model for mainframe modernization, show promising results in tasks such as COBOL summarization and defect detection [27], underscoring AI's potential in addressing legacy system challenges.

*Symbolic Execution and Constraint Solving.* Symbolic execution is a program analysis technique used to verify whether certain properties in a program, such as variable values or attributes, satisfy specific conditions [17]. In our approach, symbolic execution generates logical formulae that represent the conditions required to reproduce execution paths in the target program. These formulae are expressed in first-order predicate logic (FOPL), with variables in the formulae directly corresponding to those in the target program. To resolve these logical formulae, we leverage constraint solving, a form of automated theorem proving for FOPL. State-of-the-art SMT (Satisfiability Modulo Theories) solvers, such as Z3 [28][11], CVC5 [2], and Yices [10], are employed to handle the generated constraints effectively. These solvers have been instrumental in advancing symbolic execution by enabling precise and efficient validation of program behaviors. Research combining symbolic execution and constraint solving has shown promise in diverse areas. Notable examples include Sasaki et al.[34], Zhang et al.[44], and Sen et al. [37], which demonstrate applications ranging from test data generation to automated debugging. Despite these advancements, challenges such as scalability and handling enterprise-specific constructs remain key areas for future exploration.

*Mocking.* There are several mocking frameworks and tools for Java such as Mockito[7], EasyMock[3], WireMock[9], JMockit[6] etc. for executing Java programs. However, existing Junit generation frameworks do not generate resource mockings.

*Differential Testing.* To the best of our knowledge, we have not come across any product, tool or research that can use automatically generated unit tests to validate the equivalence of two programs across programming languages and platforms/middleware. Some early work in validating code improvement related transformation for embedded application uses equivalence check on program representation for small changes [42].

In [45], authors present a unit testing based solution to filter out invalid translations. They leveraged automatic unit test generating tools such as Evosuite (Java) and shown promising results for Java to Python and Python to C++ transformations. They do not consider calls to external resources.

There exist several tools and frameworks that can be used to solve parts of the overall problem our system tackles. Cobol-Check [1] works as a unit testing framework like JUnit for Java and generates a new COBOL program from test case and the original program which can be compiled and executed. This framework expects external resources to be setup which is not a case in a dev environment.

Test data generation for COBOL using symbolic execution has been attempted in [34], [31], [8] etc but they do not handle enterprise COBOL constructs like CICS, IBM, Db2 etc. A survey of Symbolic Execution based techniques can be found in [17].

## 5 CONCLUSION

We present a validation framework that can be used to automatically check the semantic equivalence between a COBOL program and the automatically transformed Java program. We use automatic unit test generation for the COBOL program using symbolic execution and constraint solver techniques. This framework aims at validating enterprise applications. Therefore, we not only support enterprise COBOL programming language but also the various resource access-related statements such as SQL Databases (eg IBM DB2, IMS), Transaction framework (eg, CICS, IMS) and file I/O. We use a resource input and output mechanism to solve the problem of external data access. The generated unit tests are then executed in the actual IBM Z mainframe environment by automatically generating JCLs for running the tests as jobs on the mainframe. We use a debugger script-based approach to handle resource access-related statements. Tests execution for the COBOL program generates both program output and also the resource output. These along with variable mapping (which is provided by the transformation) are used to automatically generate unit tests for translated Java programs. The output values from the mainframe are used to generate assertions in JUnit programs for testing Java.

This work is part of IBM Watsonx Code Assistant for Z (WCA4Z)[5], which includes a Validation Assistant tool[32], available as a Visual Studio Code extension. A demo video is available at [12], and program inputs on GitHub [13]. Several customer trials of WCA4Z have been conducted, but due to confidentiality, we cannot provide details. However, insights from these trials, such as the development of the extensible resource mapper and COBOL-Java mocking features, were key outcomes of client scenarios.

We show using experimental results the effectiveness of COBOL test generation by looking at the branch coverage. The Java test generation includes an additional metric which is the number of assertions generated to validate the equivalence of the COBOL and Java programs by comparing their outputs. We have implemented our validation framework as a VSCode extension which can be set up in the same VS Code IDE where the COBOL to Java transformation system is set. Various services needed to run the validation pipeline, and the mainframe system are automatically accessed by the VS Code extensions seamlessly once the URIs and credentials are configured.